\documentclass[]{emulateapj}

\begin{document}

\shorttitle{RAPTOR observations of the early optical afterglow emission from GRB 050319}
\shortauthors{Wo\'zniak et al.}

\title{RAPTOR observations of the early optical afterglow from GRB 050319}

\author{
P. R. Wo\'zniak,
W. T. Vestrand,
J. A. Wren,
R. R. White,
S. M. Evans,
and D. Casperson
}

\affil{Los Alamos National Laboratory, MS-D466, Los Alamos, NM 87545 \\
email: (wozniak, vestrand, jwren, rwhite, sevans, dcasperson)@lanl.gov}

\vspace{0.5cm}

\begin{abstract}

The RAPid Telescopes for Optical Response (RAPTOR) system at Los Alamos National Laboratory observed
GRB 050319 starting 25.4 seconds after $\gamma$-ray emission triggered the Burst Alert Telescope
(BAT) on-board the Swift satellite.  Our well sampled light curve of the early optical afterglow is composed
of 32 points (derived from 70 exposures) that measure the flux decay during the first hour after the GRB.
The GRB 050319 light curve measured by RAPTOR can be described as a relatively gradual flux decline
(power-law index $\alpha = -0.37$) with a transition, at about $\sim 400$ s after the GRB,
to a faster flux decay ($\alpha = -0.91$).  The addition of other available measurements
to the RAPTOR light curve suggests that another emission component emerged after $\sim 10^4$ s.
We hypothesize that the early afterglow emission is powered by extended energy injection or delayed
reverse shock emission followed by the emergence of forward shock emission.

\end{abstract}

\vspace{2mm}


\section{Introduction}
\label{sec:intro}

Recent years have brought interesting developments in the domain of observations of the early
optical emission from Gamma-Ray Bursts (GRBs).  Several groups now have the routine capability to respond
to GRB triggers in real time using rapidly slewing robotic instruments (e.g. \citealt{ake03,blo04,boe04,cov04,per04,ves02}).
Despite much effort in this area, so far only a handful of GRBs have been detected within the first minutes after
the onset of the $\gamma$-ray emission, namely GRBs 990123, 021004, 021211, 030418, 041219a, 050319 and 050401
(e.g. \citealt{ake99,fox03b,woz02,li03,ryk04,ves05,ryk05a},b).  Even fewer events have good S/N and coverage.

The discovery of the near infra-red transient from GRB 041219a (\citealt{bla04,bla05}) and its parallel
detection in the optical band (\citealt{wre04}) expanded the list of known GRB properties.  The RAPTOR
(Rapid Telescopes for Optical Response; \citealt{ves02}) optical light curve of GRB 041219a
(\citealt{ves05}) overlaps with the $\gamma$-ray emission by an unprecedented $\sim 6.4$ minutes. \cite{ves05} discovered
a qualitatively new component of the early optical emission from GRBs, and presented evidence for
internal shocks (\citealt{mes99}) as the emission mechanism.  The presence of the new component was
established on purely empirical basis by its distinct close correlation with strongly time-varying
$\gamma$-ray flux.

The updated taxonomy for GRB-related optical transient (OT) emission proposed by \cite{ves05} comprises:
(1) Prompt optical emission contemporaneous with and consistent with a constant flux ratio to $\gamma$-rays
(the ratio is $\sim1.2 \times 10^{-5}$ in GRB 041219a (\citealt{ves05}));
(2) Early afterglow emission that may start during the $\gamma$-ray emission and lasts for several seconds to minutes
(uncorrelated with $\gamma$-rays and typically brighter than the prompt component; e.g. GRBs 990123 and 021211); and
(3) Late afterglow emission that emerges after the fading early afterglow and can persist for hours to many days
(e.g. \citealt{fox03a}).  The current theoretical framework offers, correspondingly, internal shocks
(\citealt{mes99}), the reverse shock (\citealt{sar99,pan04}) and the external shock (\citealt{mes97,sar98})
phenomena for a possible explanation of the observed properties.

In this letter we present a comprehensive light curve of the early optical afterglow
emission from GRB 050319 starting at 35 s after the GRB trigger.


\section{Observations}
\label{sec:data}

On 2005, March 19, 09:31:18.4 UT (trigger time; hereafter $t = 0$), the Burst Alert Telescope (BAT) instrument of the Swift satellite
(\citealt{geh04}) detected GRB 050319, a single-peak event with fast raise and exponential decay lasting $T_{90} \sim 10$ s
(\citealt{kri05a,kri05b}).
The 15--350 keV fluence, the peak flux and the photon index of the time-averaged spectrum were subsequently measured to be, respectively,
$8 \times10^{-7}$ erg cm$^{-2}$, 1.7 ph cm$^{-2}$ s$^{-1}$ and 2.2$\pm$0.2 (\citealt{kri05b}). The on-board location
(\citealt{kri05a}) was distributed in near-real time through the GRB Coordinates Network (GCN) at 09:31:36.0 UT, $t = 17.6$ s.

Both the RAPTOR-S telescope and the RAPTOR-AB array responded to the alert.
RAPTOR-S is a fully autonomous robotic telescope with
0.4-m aperture and typical operating focal ratio f/5. It is equipped with a 1k $\times$ 1k pixel CCD camera employing a back-illuminated
Marconi CCD47-10 chip with 13 $\mu$ pixels. For technical details on RAPTOR-A and B see \cite{ves02}.

\begin{figure}
\vspace{9cm}
\includegraphics{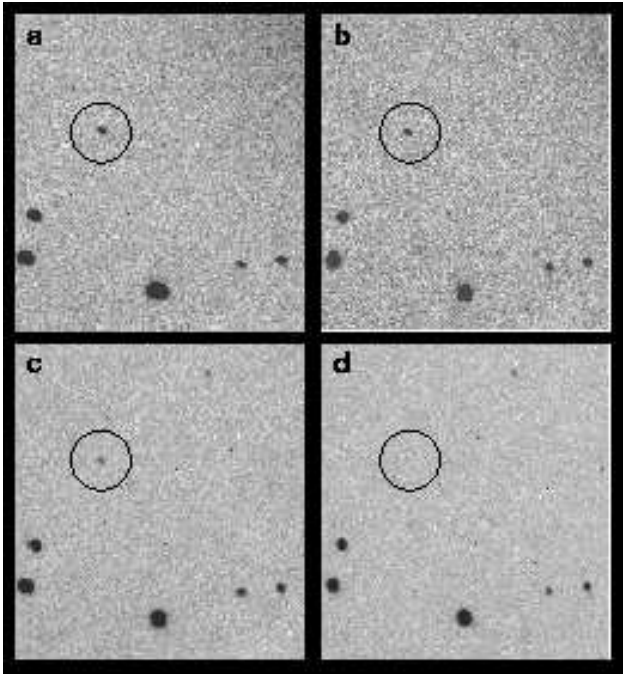}
\figcaption[]{\label{fig:frames}Examples of single RAPTOR-S exposures of GRB 050319, the OT discovered by \cite{ryk05a} is circled.
RAPTOR-S detected the OT with high S/N in the early 10-second frames (a, b), and followed its gradual decay down to the magnitude
threshold of the 30-second frames (c, d) over the next 40--50 minutes. The size of the shown area is roughly 4.5\arcmin $\times$ 4.5\arcmin
with North up and East at the left.}
\end{figure}

RAPTOR-S was on target at 09:31:53.7 UT, $t = 35.3$ s. The rapid response sequence for RAPTOR-S consists of ten 10-second images followed
by sixty 30-second images, a total of $\sim 50$ minutes of coverage (including the 15-second intervals between exposures used primarily
for readout).  A candidate OT at $\alpha$=10:16:47.9, $\delta$=+43:32:54.5 (J2000) was rapidly identified by \cite{ryk05a} within a half hour.
The OT was later confirmed by \cite{yos05}, and the absorption red-shift $z = 3.24$ was reported by \cite{fyn05}.
Initial analysis of the RAPTOR-S images (Fig.~\ref{fig:frames}) showed that the OT was detected at high S/N in 
early exposures and gradually faded below the magnitude limit.  Unfortunately, the observing conditions at the RAPTOR-S site during response
were variable and clouds obscured the field of view between $t = 1480$ and 2440 s.

RAPTOR-B instrument responded slightly faster.  Although none of those images is a detection, including the first 10-second frame starting
at $t = 25.44$ s, the corresponding magnitude limit for OT is of some value (\S\,\ref{sec:results}).
RAPTOR-B and S are separated by about 37 km and have independent weather.

\section{Photometry}
\label{sec:redux}

After standard corrections for bias, dark current and flat field responses, all frames underwent a 2$\times$2 pixel binning.
The binning was applied in order to bring the sampling of the stellar images to about the critical Nyquist value.
Additionally, it increased S/N per pixel and made the point-spread function (PSF) nearly circular.


We rejected 29 images taken between $t = 1480$ and 2440 s, when transparency was very poor due to passing clouds.
In even later images the OT detections are marginal. We decided, therefore, to form two mean averages
of 5 and 6 frames out of 11 images taken after $t = 2440$ s.  A high S/N reference image was prepared
by mean stacking twenty 30-second frames. All object centroids (including OT) were determined using the reference image.
For the purpose of averaging and photometric analysis, all frames were resampled to a common pixel grid using a bi-cubic
spline interpolator and a linear coordinate transformation with the $\sim$0.08 pixel accuracy (r.m.s.) based on
the positions $\sim30$ high S/N field stars.  For image processing we used a custom Difference Image Analysis software
(\citealt{woz00}).

PSF-weighted photometry within a 4-pixel radius was performed assuming a fixed centroid (from reference image) and without
variance weighting.  This general technique, used in Sloan Digital Sky Survey (Lupton 2005, in preparation), hedges against a secondary
nonlinearity between the bright and faint ends of the flux scale.  It ensures that the much flatter variance profile of the background dominated
objects cannot propagate the systematic uncertainties from the PSF shape to the photometric offsets. We assumed a Gaussian
PSF with FWHM = 6\arcsec\, (2.44 pixel, binned).  The flux scale with about 3.8\% internal consistency was established using 11 high S/N stars
in the vicinity of the OT. The calibration to standard $R$ magnitudes was based on measurements of 22 USNO-A2.0 stars in the magnitude range
$R2$ = 12.5--18.5.  Residuals with respect to the best constant magnitude offset were random over the full flux range (good linearity)
with r.m.s. scatter of 0.09 mag outside the photon noise dominated region.  Our unfiltered optical band has an effective
wavelength close to that of the standard $R$ band, but it has a larger width.  For lack of the instrumental color information, we assumed
that all objects have the color of a mean comparison star, i.e. $(B-R)$ = 1.25 and $(R-I)$ = 0.73, according to USNO-A2.0 catalog.
The fact that colors of the early GRB afterglows and their temporal evolution remain poorly constrained is a source of major uncertainty
in transformations of broad band photometry (compare \S\,{\ref{sec:results}).  


The early RAPTOR-B frame was analyzed using the same techniques as applied to RAPTOR-S data.
The actual limit was calculated by performing a fixed centroid PSF photometry at numerous random locations near the nominal
OT position, taking the r.m.s. of the measured flux and converting to magnitudes.


\section{Results}
\label{sec:results}

The final RAPTOR photometry of the early optical afterglow of GRB 050319 expressed on the $R$-band scale is given
in Table~\ref{tab:data}. Fig.~\ref{fig:lc} plots the light curve along with our model fits.
\cite{qui05} found an acceptable fit to an unfiltered optical light curve from ROTSE-IIIb telescope using a single
power-law model with $\alpha = -0.59 \pm 0.05$.  For the RAPTOR data the best fitting single power-law model
has index $\alpha = -0.55 \pm 0.02$, however it yields an unacceptable $\chi^2/d.o.f. = 9.20$.  A visual inspection of the RAPTOR
measurements suggests a shallow flux decay at early times and significant steepening after $\sim 400$ s.
In fact, the residuals with respect to the best fitting single power-law model are systematic and indicate
a steepening trend.  To test that hypothesis we fitted a broken power-law model and obtained a reasonably good fit
($\chi^2$/d.o.f. = 2.91) with $\alpha_1 = -0.38 \pm 0.03$, $\alpha_2 = -0.91 \pm 0.06$ and the break time
$t_{\rm br} = 462 \pm 55$ s.  It should be noted that instantaneous scale breaking in this direction may be hard to explain
physically.  Nevertheless, we find the model useful for investigating possible changes in the light curve slope.

The residuals with respect to the best fitting broken power-law model appear flat, however the reduced $\chi^2 = 2.91$ (81.42/28 d.o.f.)
is still formally unacceptable.  For some measurements the deviations from the best fit model
are well in excess of the error estimates, in particular there are several strong outliers right near the fitted time of
the break.  There are also a few points with fluxes significantly larger than the model prediction right before passing clouds
covered the field of view.  The additional photometric scatter could be related to variable observing conditions
and is discussed in more detail in \S\,\ref{sec:discussion}.

To establish the significance of the break, we fitted a series of broken power-law models with a range of fixed
break times.  We found that the minimum of the $\chi^2$ surface is not very well constrained and the actual 68\% confidence
interval for the break time may be closer to $\pm$100 s than to the formal parabolic error bar.
Our conclusion is that the RAPTOR data indicate a significant steepening in the flux decay of GRB 050319 within the first hour
after the $\gamma$-ray trigger.  However, despite the appearance of sharp break in the light curve near $\sim400$ s, our data are fully
consistent with a gradual increase of the slope, possibly with additional small scale photometric variations.

\begin{figure}
\vspace{10cm}
\includegraphics{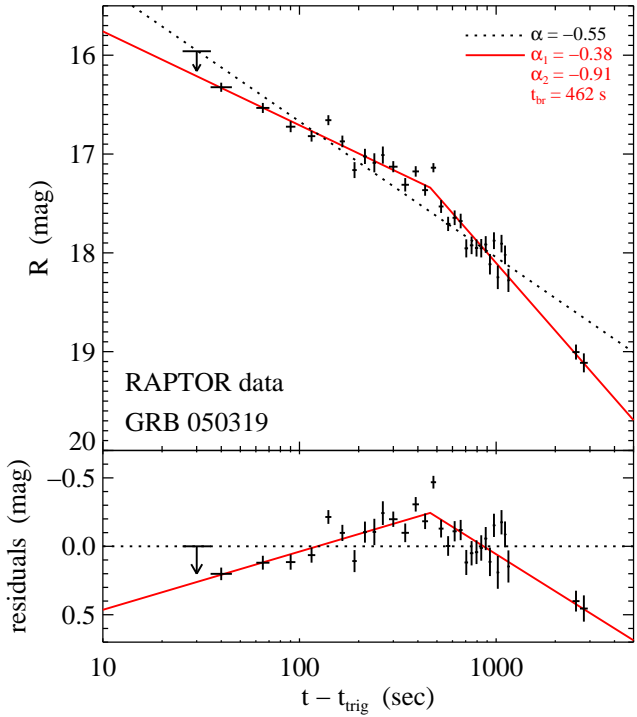}
\figcaption[]{\label{fig:lc}RAPTOR-S optical light curve of GRB 050319 (top) and photometric
residuals with respect to the reference model (bottom). Our reference model, i.e. the best fitting single
power-law (black dotted line), produces systematic residuals. We obtain a much better fit with
a broken power-law model (red solid line), which indicates significant steepening around $\sim 400$ s.}
\end{figure}

\begin{deluxetable}{rrccc}
\tablewidth{10cm}
\tablecaption{\label{tab:data}{RAPTOR photometry of GRB 050319.\tablenotemark{a}}}
\tablehead{
\colhead{$t_{\rm start}$} &
\colhead{$t_{\rm end}$} &
\colhead{$\Delta t_{\rm exp}$} &
\colhead{$R$} &
\colhead{$\sigma$} \\
\colhead{\makebox[1.7cm]{(s)}} &
\colhead{\makebox[1.7cm]{(s)}} &
\colhead{\makebox[1.0cm]{(s)}} &
\colhead{\makebox[1.0cm]{(mag)}} &
\colhead{\makebox[1.0cm]{(mag)}}
}
\startdata
\dotfill     25.4 & \dotfill     35.4 &    10 &  $>$15.960~~ & \nodata  \\
\dotfill     35.3 & \dotfill     45.3 &    10 &     16.323 &    0.046  \\
\dotfill     60.3 & \dotfill     70.3 &    10 &     16.532 &    0.050  \\
\dotfill     85.3 & \dotfill     95.3 &    10 &     16.722 &    0.056  \\
\dotfill    110.3 & \dotfill    120.3 &    10 &     16.818 &    0.055  \\
\dotfill    135.3 & \dotfill    145.3 &    10 &     16.656 &    0.050  \\
\dotfill    160.3 & \dotfill    170.3 &    10 &     16.870 &    0.059  \\
\dotfill    185.5 & \dotfill    195.5 &    10 &     17.160 &    0.080  \\
\dotfill    210.3 & \dotfill    220.3 &    10 &     17.021 &    0.077  \\
\dotfill    235.3 & \dotfill    245.3 &    10 &     17.084 &    0.096  \\
\dotfill    260.5 & \dotfill    270.5 &    10 &     17.007 &    0.087  \\
\dotfill    285.3 & \dotfill    315.3 &    30 &     17.127 &    0.056  \\
\dotfill    330.3 & \dotfill    360.3 &    30 &     17.309 &    0.068  \\
\dotfill    375.3 & \dotfill    405.3 &    30 &     17.175 &    0.053  \\
\dotfill    420.5 & \dotfill    450.5 &    30 &     17.364 &    0.057  \\
\dotfill    465.2 & \dotfill    495.2 &    30 &     17.138 &    0.047  \\
\dotfill    510.2 & \dotfill    540.2 &    30 &     17.529 &    0.065  \\
\dotfill    555.2 & \dotfill    585.2 &    30 &     17.705 &    0.072  \\
\dotfill    600.2 & \dotfill    630.2 &    30 &     17.642 &    0.074  \\
\dotfill    645.2 & \dotfill    675.2 &    30 &     17.675 &    0.074  \\
\dotfill    690.2 & \dotfill    720.2 &    30 &     17.951 &    0.091  \\
\dotfill    735.4 & \dotfill    765.4 &    30 &     17.920 &    0.089  \\
\dotfill    780.2 & \dotfill    810.2 &    30 &     17.948 &    0.087  \\
\dotfill    825.2 & \dotfill    855.2 &    30 &     17.949 &    0.093  \\
\dotfill    870.2 & \dotfill    900.2 &    30 &     17.912 &    0.085  \\
\dotfill    915.2 & \dotfill    945.2 &    30 &     18.110 &    0.104  \\
\dotfill    960.4 & \dotfill    990.4 &    30 &     17.874 &    0.085  \\
\dotfill   1005.2 & \dotfill   1035.2 &    30 &     18.241 &    0.120  \\
\dotfill   1050.2 & \dotfill   1080.2 &    30 &     17.903 &    0.089  \\
\dotfill   1095.2 & \dotfill   1125.2 &    30 &     18.016 &    0.094  \\
\dotfill   1140.4 & \dotfill   1170.4 &    30 &     18.272 &    0.118  \\
\dotfill   2445.0 & \dotfill   2654.9 &   150 &     19.001 &    0.076  \\
\dotfill   2669.9 & \dotfill   2924.9 &   180 &     19.109 &    0.096  \\
\enddata

\tablenotetext{a}{All measurements were obtained with the RAPTOR-S instrument except the limit
at $t = 25$ s recorded by RAPTOR-B.  Our unfiltered magnitudes were transformed to $R$-band scale
using USNO-A2.0 catalog, and were not corrected for extinction (Galactic $E(B-V)$ reddening is only 0.01 mag;
\citealt{sch98}).  The last two images are stacks of five and six 30-second frames.
For those images the effective exposure time accounts for readout breaks and is shorter than the difference between
the end and start times ($\Delta t_{\rm exp} < t_{\rm end} - t_{\rm start}$).}

\end{deluxetable}

\section{Discussion}
\label{sec:discussion}

\begin{figure}
\vspace{10.2cm}
\includegraphics{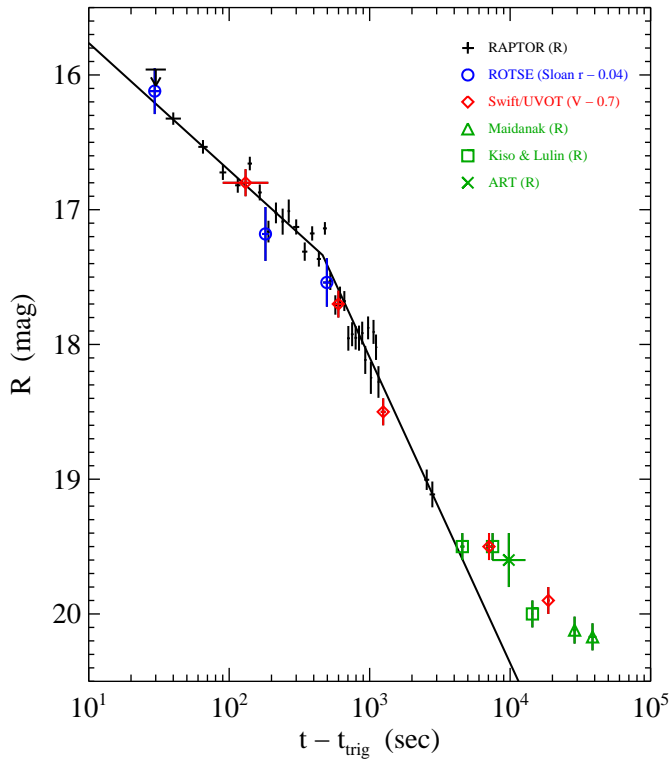}
\figcaption[]{\label{fig:comb_lc}Comparison of the RAPTOR optical light curve of GRB 050319
with measurements from other instruments. The black points are RAPTOR measurements,
while the data plotted in color are those obtained by other instruments: ROTSE (\citealt{qui05}),
UVOT (\citealt{boy05}), Maidanak 1.5m (\citealt{sha05}), Kiso \& Lulin 1m (\citealt{yos05}),
and ART 14-inch (\citealt{tor05}). The line shows our best fit broken power-law model.}
\end{figure}

To test the robustness of the results of \S\,\ref{sec:results} we reanalyzed the data using three other photometric tools to extract object fluxes:
(1) traditional aperture photometry; (2) kernel matched difference image photometry (\citealt{woz00}); (3) a standard PSF package DoPHOT
(\citealt{sch93}), and obtained essentially identical light curves.  While the steepening of the light curve and most wiggles were always present,
the precise origin of photometric outliers in Fig.~\ref{fig:lc} still escapes explanation.  Some comparison stars also show the wiggles and some are
fully consistent with the photon noise estimate.  \cite{li03} noticed similar discrepancies in their light curve of GRB 021211 and suggested that
color induced systematics could be the cause.  The intrinsic variability of the OT color or even a stationary color difference between the OT and comparison
stars may generate systematic offsets in photometry.  Given that the unfiltered spectral band is subject to a red atmospheric cutoff and weather variations
during RAPTOR response, it is entirely possible that the outliers come from a systematic effect yet to be found.  This experience underscores
the importance of simultaneous color measurements of the early GRB afterglows using standard filters.

In Fig.~\ref{fig:comb_lc} we plot other OT measurements available at this time for comparisons with the RAPTOR light curve.  The three points from
ROTSE collaboration (\citealt{qui05}) were shifted by $-0.04$ mag to reflect the median difference between SDSS $r$ and USNO-A2.0 $R$ magnitudes
for our comparison stars.  The $V$-band points from Swift/UVOT (\citealt{boy05}) are plotted 0.7 mag brighter than actual values.  All measurements
reported in standard $R$ band were taken at face value, since any finer issues with calibration to different catalogs should wait until the final revised
photometry is available.

Measurements by other experiments agree with the shape of the RAPTOR light curve.  While more sparsely sampled, the $V$-band
light curve observed by the Swift/UVOT (\citealt{boy05}) also shows a faster flux decay after $\sim 400$ s.  Further, the measurement at $t = 1.27$ hours
by \cite{yos05} is consistent with the extrapolated value predicted by the RAPTOR measurements.  At times beyond $\sim 1.3$ hours, the published
data from UVOT and other instruments show a transition back to a more gradual flux decay rate.  The two breaks in the flux decay rate are visible
in both the $R$-band and the $V$-band light curves.  The close tracking between the light curves in both filters implies a constant $(V-R)$ color
during the first day of the flux evolution.  The zero point for the UVOT measurements is still not well established (\citealt{boy05}),
but taken at face value the reported measurements yield a color $(V-R) = 0.7$ for the OT counterpart of GRB 050319.

While the sample size is still small, one can already start to explore the morphology of GRB early afterglow light curves.
The early afterglow behaviors of GRBs 990123 (\citealt{ake99}) and 021211 (\citealt{li03}) were very similar, with both OTs showing
a steeper initial decline (power-law index $\alpha \simeq -1.8$) and the emergence of a shallower component ($\alpha \simeq -0.9$)
after $\sim$10 minutes. On the other hand, GRBs 021004 (\citealt{fox03b}) and 030418 (\citealt{ryk04}) showed shallower initial decline
(or even rise) with $\alpha > -0.6$ and then gradual steepening (to about $\alpha < -1.0$) on time-scales of $\sim 10^3$ s or longer.
The measurements we reported here for GRB 050319, starting from $\alpha_1 = -0.38$ and evolving to $\alpha_2 = -0.91$ after $\sim 400$ s,
place its early afterglow properties in the latter group.

In the context of the standard fireball model, the shape of the optical afterglow light curve is determined by the nature of the interaction
between the relativistic ejecta and the external medium. The relative importance and timing of the reversed and forward shock components,
which depend on properties like the density profile in the external medium and the strength of the magnetic field of the fire-ball,
are reflected in the rates of flux evolution and the break times in the predicted light curve (e.g. \citealt{sar99,mes02,zha04}).
The morphology of the GRB 990123 and 021211 light curves, with the break to shallower decay, is usually attributed to the transition
from the dominance of the reverse shock generated emission to forward shock generated emission (e.g. \citealt{li03}).  For the more gradually
decaying early afterglows, the interpretation is less clear.  The gradually declining component could be associated with delayed reverse shock
emission (\citealt{ves05}) or an energy injection that continues well beyond the duration of the initial explosion (\citealt{fox03b}).
The emergence of the additional component after $\sim 10^4$ s in GRB 050319 could then be understood as the emergence of the forward shock
emission. The prompt Swift localizations and rapid robotic followup are just starting to reveal the richness and complexity of the GRB
afterglow phenomenology and, when combined with the models, will help to constrain the basic physical parameters of these cataclysmic
explosions.

\acknowledgements

This research was performed as part of the Thinking Telescopes project supported by the Laboratory Directed Research
and Development (LDRD) program at LANL. PW was supported by the Oppenheimer Fellowship.

\end{document}